# *Biaxial nematics: symmetries, order domains and field-induced phase transitions*[#]

S. D. Peroukidis, P. K. Karahaliou[*], A. G. Vanakaras and D. J. Photinos

*Department of Materials Science, University of Patras, Patras 26504, Greece*

**Abstract**

We study the symmetry and the spatial uniformity of orientational order of the biaxial nematic phase in the light of recent experimental observations of phase biaxiality in thermotropic bent-core and calamitic-tetramer nematics. We present evidence supporting monoclinic symmetry, instead of the usually assumed orthorhombic. We describe the use of deuterium NMR to differentiate between the possible symmetries. We present the spatial aspects of biaxial order in the context of the cluster model, wherein macroscopic biaxiality can result from the field-induced alignment of biaxial and possibly polar domains. We discuss the implications of different symmetries, in conjunction with the microdomain structure of the biaxial phase, on the alignment of biaxial nematics and on the measurements of biaxial order.

## 1. Introduction

Biaxial nematics are still very rare compounds, four decades after their initial theoretical prediction *(1)* and despite numerous subsequent confirmations and extensions of the prediction by molecular theory and computer simulations for a variety of model systems *(2)*, in parallel with extensive synthetic and characterization efforts encompassing a broad range of molecular architectures *(3-12)*. Interestingly, the few known instances of biaxial nematics belong to quite diverse physico-chemical categories of liquid crystals: The first biaxial nematic to be realized experimentally *(3)* was lyotropic. A decade later, phase biaxiality was reported for side-chain liquid crystal polymers, *(4)*. Low molar mass thermotropic biaxial nematics were achieved more recently, first in bent-core *(9)* mesogens and shortly afterwards in radial tetramers of laterally tethered calamitic mesogens *(10)*. Lastly, recent synthetic *(6-7, 12)* and computer simulation *(13)* results have renewed optimism for the achievement of the first biaxial rod-plate thermotropic nematic mixture.

In addition to their intrinsic scientific interest, biaxial nematics, particularly of the thermotropic low-molar mass type, are of potential importance to the display technology. This is mainly due to the expected speed advantage of the electro-optic response of the transverse ("biaxial") optical axes over the response of the conventional uniaxial nematics presently used in LC displays, which involves the reorientation of the primary director **n**. While the first confirmations of the speed advantage of biaxial nematics have been obtained, both experimentally *(14)* and from computer simulations *(15)*, a number of fundamental materials-related issues are to be successfully addressed before this advantage can be actually harvested in competitive display applications. In this respect, an immediate objective is to systematically engineer room-temperature low molar mass biaxial nematics. Another major issue is to control surface-alignment of the primary and the transverse optical axes and their selective addressability by applied electric fields. Furthermore, quantitative control over basic materials properties is essential. Thus the ability to reorient the transverse axes using fields of reasonable magnitude requires substantial values of the transverse dielectric anisotropy. Similarly, a substantial transverse optical biaxiality is required if the optical effect of reorienting the transverse axes is to be obtained from sufficiently thin layers of the biaxial material. A closely related issue is that of the elastic constants: In the operating modes of uniaxial nematic display devices, the elastic constants provide the restoring mechanism when the applied field is switched off. The biaxial elastic constants are normally estimated to be considerably weaker than the uniaxial ones, signaling possible difficulties with the restoring mechanism for biaxial modes. Enhancing the relevant elastic constants without deteriorating the speed advantage of the biaxial modes is one of the advanced goals of biaxial nematic molecular design. On the other hand, the difficulties posed by the weak biaxial elasticity could in principle be bypassed in biaxial nematics that exhibit spontaneous transverse electric polarization (polar biaxial nematics), in which case the switching as well as the restoring mechanism could be driven by the electric field.

---

[#] This paper is based on an invited lecture presented at the 22[nd] International Liquid Crystal Conference (Jeju, Korea, 2008).

[*]Corresponding Author. Email: pkara@upatras.gr

Undoubtedly, the design and synthesis of materials to meet the above requirements, constitute one of the major challenges in contemporary liquid crystal science. The recent experimental observation of phase biaxiality in thermotropic nematic liquid crystals *(9-10)*, has stimulated and intensified research efforts in that direction. To date, a very significant outcome of these efforts has been a number of findings whose consistent interpretation could broaden substantially the previous perceptions of the biaxial nematic phase. In particular, there are strong indications that (i) the symmetry of the biaxial nematic phase is not necessarily restricted to the $D_{2h}$ point group, which was assumed in the original theoretical predictions *(1-2)* and later applied successfully for the analysis of the first experimental observations of phase biaxiality in lyotropic *(3)* and polymer *(4, 16)* nematics, and (ii) the biaxial phase may exhibit a hierarchical domain structure which differs qualitatively from the typical domain structure of uniaxial nematics and which could underlie the appearance of substantial field-induced biaxiality *(17)* at relatively low field strengths as well as the appearance of chiral domains in molecularly achiral bent-core compounds *(18)* and the appearance of spontaneously polar domains *(19)*. In this broader perspective of the biaxial phase, new insights can be gained towards answering long standing questions concerning the experimentally measurable signatures of nematic phase biaxiality, the related question of whether the measured macroscopic biaxial order is in some cases induced by the conditions of the measurement itself and, eventually, the question of why thermotropic biaxial nematics are so hard to obtain experimentally. The purpose of this work is to present the symmetry and domain-structure implications of currently available experimental observations and to incorporate them into a model that embodies this broader view on the biaxial nematic phase. The symmetry aspects of the biaxial nematic phase are considered in Section 2. The local structures and their influence on the transitions from uniaxial to biaxial states are presented in Section 3. The implications of the symmetry and domain structure on the alignment of biaxial nematics are discussed in the concluding section.

**2. The symmetries of the biaxial nematic phase.**

The first theoretical predictions *(1)* and essentially all the molecular theory and computer simulation work that followed on biaxial nematics *(2)*, referred to a phase of $D_{2h}$ symmetry, with the three mutually orthogonal symmetry axes defining the directors, **n, l, m,** of the phase. Furthermore, the first experimental observations of phase biaxiality, from *NMR* studies on a lyotropic liquid crystal *(3)*, where interpreted consistently on the basis of the $D_{2h}$ symmetry. Likewise, the results of the first optical studies *(4)* and of the subsequent *NMR* observations *(16)* of phase biaxiality in the nematic phase of side-chain liquid crystal polymers, where consistent with the $D_{2h}$ symmetry. The situation is, however, different for the presently available observations of biaxial orientational order in thermotropic nematics. As detailed in ref *(20)*, the experimental results involving measurements on aligned samples, although routinely analysed assuming a $D_{2h}$ symmetry, do not particularly support this symmetry and in some cases appear to contradict it. Thus, the *NMR* measurements on bent-core nematics *(9)* using deutriated probe solutes indicate the presence of biaxial order but are not detailed enough to single out a particular symmetry for the observed phase biaxiality. On the other hand, the biaxial order in calamitic tetramer nematics, reported first from *IR* spectroscopy studies on aligned samples *(10)* and subsequently confirmed by *NMR* measurements using deuteriated probe solutes *(21)*, when analyzed on the assumption of $D_{2h}$ symmetry leads to inconsistently large values of biaxial order parameters and to mutually conflicting inferences regarding the molecular attributes underlying phase biaxiality *(20)*. In addition to these difficulties, *X*-ray diffraction studies in the nematic phase of bent-core compounds *(22-23)*, as well as in the nematic phase of the side-on monomers which form the calamitic tetramer compounds *(24-25)*, show the presence of local biaxial order that is characteristic of tilted-layer domain structures, thus favouring a monoclinic rather than orthorhombic symmetry for the biaxial phase that is formed when these domain structures become macroscopically ordered in the transverse direction *(17)*.

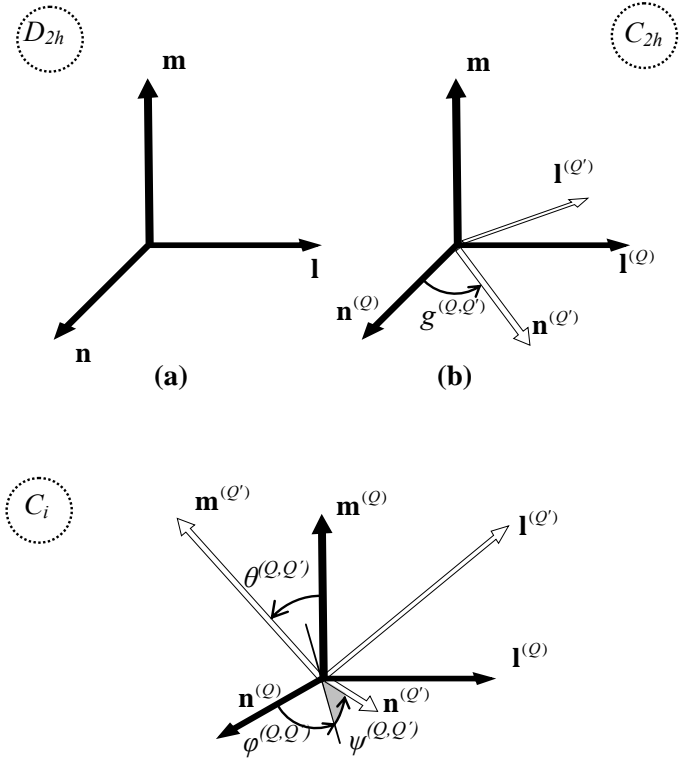

**Figure 1.** Principal axes and directors for apolar, achiral biaxial phases of different symmetries. (a) Orthorhombic ($D_{2h}$) phase. The directors $\mathbf{n}, \mathbf{l}, \mathbf{m}$ are identified with the twofold symmetry axes of the phase. (b) Monoclinic ($C_{2h}$) phase, in this instance with the twofold symmetry axis of the phase defining the single director $\mathbf{m}$. (c) Triclinic phase ($C_i$), with the principal frames pertaining to two distinct tensor properties $O, O'$ of the medium.

Theoretically, the orthorhombic $D_{2h}$ symmetry is only one of several point group symmetries that are possible for the biaxial nematic phase *(26-27)*. For apolar, achiral phases, to which we restrict our attention in this section, the possible point group symmetries of a biaxial nematic are three (20): the orthorhombic $D_{2h}$, the monoclinic $C_{2h}$ and the triclinic $C_i$. Higher up in symmetry, the uniaxial nematic phase belongs to the $D_{\infty h}$ group, with the axis of full rotational symmetry (i.e. the director $\mathbf{n}$) defining the principal direction of all the macroscopic anisotropic physical properties of the medium. The anisotropy of this phase is reflected primarily on second-rank traceless tensor properties $Q_{AB}$. The director $\mathbf{n}$ is the principal axis for all such properties and is normally identified with the principal $Z$ axis of the phase, the choice of the other two orthogonal axes $X, Y$, being arbitrary. Any tensor property is then fully described by its principal value $Q_{ZZ}$ in that frame, since $Q_{XX} = Q_{YY} = -Q_{ZZ}/2$. The distinction between a uniaxial and a biaxial nematic, is that the latter has at least one tensor property $Q_{AB}$ whose principal values are all different, i.e. $Q_{XX} \neq Q_{YY} \neq Q_{ZZ}$. The specific choice of the principal axes is usually made such that $|Q_{XX}| < |Q_{YY}| < |Q_{ZZ}|$, in which case the principal values of the tensor are represented by $Q_{ZZ}$ and the biaxiality parameter $\eta^{(Q)} \equiv (Q_{XX} - Q_{YY})/Q_{ZZ}$. The distinction between biaxial phases of different symmetries lies in the relative orientations of the principal axis frames of the different tensor properties: In the orthorhombic ($D_{2h}$) phase, the three two-fold symmetry axes define a unique triplet of directors $\mathbf{n}, \mathbf{l}, \mathbf{m}$ which are the principal axes of all the macroscopic tensor properties $Q_{AB}$ of the medium (Figure 1a). The monoclinic ($C_{2h}$) phase has one unique director, coincident with the single two-fold symmetry axis of the phase, which is a common principal axis for all the second-rank tensor properties of the medium. The other two principal axes, however, are not in general common for all the tensor properties. Thus, if the symmetry axis is identified, for example, with the common director $\mathbf{m}$ (Figure 1b), the other two principal axes $\mathbf{n}^{(Q)}, \mathbf{l}^{(Q)}$ and $\mathbf{n}^{(Q')}, \mathbf{l}^{(Q')}$ of two different tensors $Q_{AB}$ and $Q'_{AB}$ will be in general rotated relative to one another by an angle $g^{(Q,Q')}$ about the symmetry axis **m.** Lastly, the triclinic phase ($C_i$) has no unique director; different tensors $Q_{AB}$, $Q'_{AB}$ have in general principal frames $\left(\mathbf{n}^{(Q)},\mathbf{l}^{(Q)},\mathbf{m}^{(Q)}\right)$, $\left(\mathbf{n}^{(Q')},\mathbf{l}^{(Q')},\mathbf{m}^{(Q')}\right)$ which differ in the directions of all three axes (Figure 1c).

According to the above, the parameters $\eta^{(Q)}, \eta^{(Q')}$..., associated with the various measurable second rank tensor properties of the medium, fully quantify biaxiality in an orthorhombic nematic phase. In a monoclinic phase, the quantification of biaxiality involves, in addition to the individual parameters $\eta^{(Q)}, \eta^{(Q')}$..., the measurable angles $g^{(Q,Q')}$ of relative rotation of the principal frames associated with different pairs of tensor properties; these angles provide the means of experimental distinction between the monoclinic symmetry ($\sin 2g^{(Q,Q')} \neq 0$ for at least one pair of tensor quantities $Q,Q'$) and the orthorhombic symmetry ($\sin 2g^{(Q,Q')} = 0$ for any pair of tensor quantities $Q,Q'$). Similarly, the triclinic symmetry involves, in addition to the individual biaxiality parameters $\eta^{(Q)}, \eta^{(Q')}$..., the three angles $\theta^{(Q,Q')}, \phi^{(Q,Q')}, \psi^{(Q,Q')}$ of relative rotations of the principal frames, which provide the experimental distinction of the triclinic from the monoclinic and orthorhombic symmetry.

Clearly, the possibility of apolar, achiral biaxial nematic phases of different symmetry, could have profound implications on the experimental identification of phase biaxiality, on its quantification and on the alignment and electro-optic properties of these materials. In the remainder of this section we will focus on the experimental differentiations between the orthorhombic and the monoclinic symmetry. The crucial differentiating quantities in this case are the relative rotation angles $g^{(Q,Q')}$ between the principal axes of different tensor quantities of the medium. There is a variety of such quantities, ranging from materials properties, such as the static dielectric tensor, the magnetic susceptibility tensor and the index of refraction ellipsoid, to molecular-site-specific tensor properties, such as the qudrupolar splittings measured by *NMR* methods *(28-29)* and the anisotropic absorbances measured in *IR* spectroscopy *(10, 30)*. It is known, however, from the study of biaxial order in the *Smectic-C* phase, which is also of $C_{2h}$ symmetry, that the differences in the orientations of the principal axes associated with the dielectric, the optical and the diamagnetic anisotropies are rather small for common calamitic compounds *(30)*. This is normally attributed to the fact that these anisotropies originate from the rod-like mesogenic core of the molecules, which, furthermore, undergo extensive rotational averaging about their long axis, thus diminishing on the macroscopic scale the relative deviations of the principal axes. On the other hand, molecular-site-specific properties, notably the motionally averaged quadrupolar interactions of deuterated molecular sites, could show substantial relative deviations of their principal axes *(31)* since distinct labelled sites of the flexible molecules can be influenced differently by the rotational averaging about the molecular axis. In this respect, the measurements of site-specific properties are advantageous over the measurements of global molecular properties. In particular, deuterium *NMR* (*2H-NMR*) spectroscopy using selectively deuterated molecules has been one of the most powerful tools in the study of nematic phase biaxiality *(3, 9, 16, 21, 32-35)*. To date, the *2H-NMR* measurements in these studies have been limited to the determination of the biaxiality parameters $\eta$ associated with the orientationally averaged quadrupolar interaction of selectively dueteriated molecular sites of the pure compounds or of probe-solutes. As we discuss in the remainder of this section, the same *2H-NMR* methods are particularly suited for the differentiation between the orthorhombic and the monoclinic symmetry in biaxial nematics.

First, the analysis of the *2H-NMR* spectra involves several second rank tensor quantities whose principal axis frames can be determined relative to the magnetic field of the spectrometer *(28-29)*; it therefore offers a multitude of options for identifying possible differences in the orientations of distinct frames. The tensor quantities involved are: (a) The diamagnetic anisotropy $\chi^m_{AB}$ of the phase. This tensor governs the orientation of the sample relative to the magnetic field. Its principal axes $X_M, Y_M, Z_M$, are specifically chosen so that $\chi^m_{X_M X_M} \leq \chi^m_{Y_M Y_M} \leq \chi^m_{Z_M Z_M}$, i.e. by identifying $Z_M$ and $X_M$ with the directions of lowest and highest magnetic energy respectively. (b) The qudrupolar interaction at each deuteriated site *i* is conveyed by a second-rank symmetric and traceless tensor $G_{AB}^{(i)}$, describing the orientational averaging of the field gradient associated with the molecular site *(20)*. The principal axes of each such tensor, $X_i, Y_i, Z_i$ are chosen so that $Z_i(X_i)$ corresponds to the largest (smallest) absolute principal value, and the tensor is represented by the principal order parameter $S^{(i)}$ and the biaxiality parameter $\eta^{(i)}$. (c) Some of the deuteriated sites may exhibit substantial anisotropic chemical shift asymmetry (*CSA*) in which case a measurable tensor $C_{AB}^{(i)}$ will be associated with each of these sites. The principal axis frame of this tensor $C_{AB}^{(i)}$ is not necessarily coincident with that of $G_{AB}^{(i)}$ (unless, of course, the phase is orthorhombic). (d) For closely positioned pairs of deuteriated sites *i,i'*, the dipolar couplings, give rise to a measurable tensor $D_{AB}^{(i,i')}$, with its own principal axis frame, distinct from those of $G_{AB}^{(i)}$ and $C_{AB}^{(i)}$.

For a monodomain sample, oriented so that the magnetic field forms the angles $\theta_M, \phi_M$ relative to the magnetic principal axes $X_M, Y_M, Z_M$ of the sample, the measurable qudrupolar frequency spectrum consists of a set of double peaks. Each doublet corresponds to a distinct deuteriated molecular site *i*, or group of equivalent sites, and presents a frequency splitting $\delta v^{(i)}$. The orientation dependence of the splittings is controlled by the components of the tensor $G_{AB}^{(i)}$ in the frame of the magnetic principal axes $X_M, Y_M, Z_M$ and is given by *(20)*:

$$\delta v^{(i)}(\theta_M,\phi_M) = \frac{3}{2}v_Q^{(i)}\left[\begin{array}{l}\left(\frac{3}{2}\cos^2\theta_M - \frac{1}{2}\right)G_{Z_MZ_M}^{(i)} + \frac{1}{2}\sin^2\theta_M \cos 2\phi_M\left(G_{X_MX_M}^{(i)} - G_{Y_MY_M}^{(i)}\right)\\ + \sin 2\theta_M \cos\phi_M G_{Z_MX_M}^{(i)} + \sin 2\theta_M \sin\phi_M G_{Z_MY_M}^{(i)} + \sin^2\theta_M \sin 2\phi_M G_{Y_MX_M}^{(i)}\end{array}\right]. \quad (1)$$

Here, $v_Q^{(i)}$ is the quadrupolar coupling constant for the deuteriated site *i*. The symmetry of the phase is reflected on the spectra through the five independent components of $G_{AB}^{(i)}$ which appear in equation (1). In what follows we consider a particular example of symmetry for which a full determination of the $G_{AB}^{(i)}$ tensor, and therefore of the phase symmetry, is possible by a technique combining the measurement of the spectrum of the aligned sample with the spectrum accumulated as the sample is spinning about an axis perpendicular to the magnetic field *(9,32,34,36-38)*. To this end, we take the biaxial nematic phase to have $C_{2h}$ symmetry or higher and to have the magnetic $X_M$ axis (i.e. the axis of maximum magnetic energy) coinciding with the twofold rotation symmetry axis of the phase. This makes the components $G_{Z_MX_M}^{(i)}, G_{Y_MX_M}^{(i)}$ in equation (1) vanish by symmetry. In this case, by analysing the spectra from the magnetically aligned and from the spinning sample one can (a) single out which of the principal axes $X_i, Y_i, Z_i$ of the $G_{AB}^{(i)}$ tensor coincides with the two-fold symmetry axis of the phase (and therefore with the principal magnetic axis $X_M$) and (b) provide the values of the primary order parameter $S^{(i)}$, the biaxiality parameter $\eta^{(i)}$ and the angle $g^{(i)}$ by which the $X_i, Y_i, Z_i$ frame is rotated relative to the $X_M, Y_M, Z_M$ frame about the common symmetry axis of the phase.

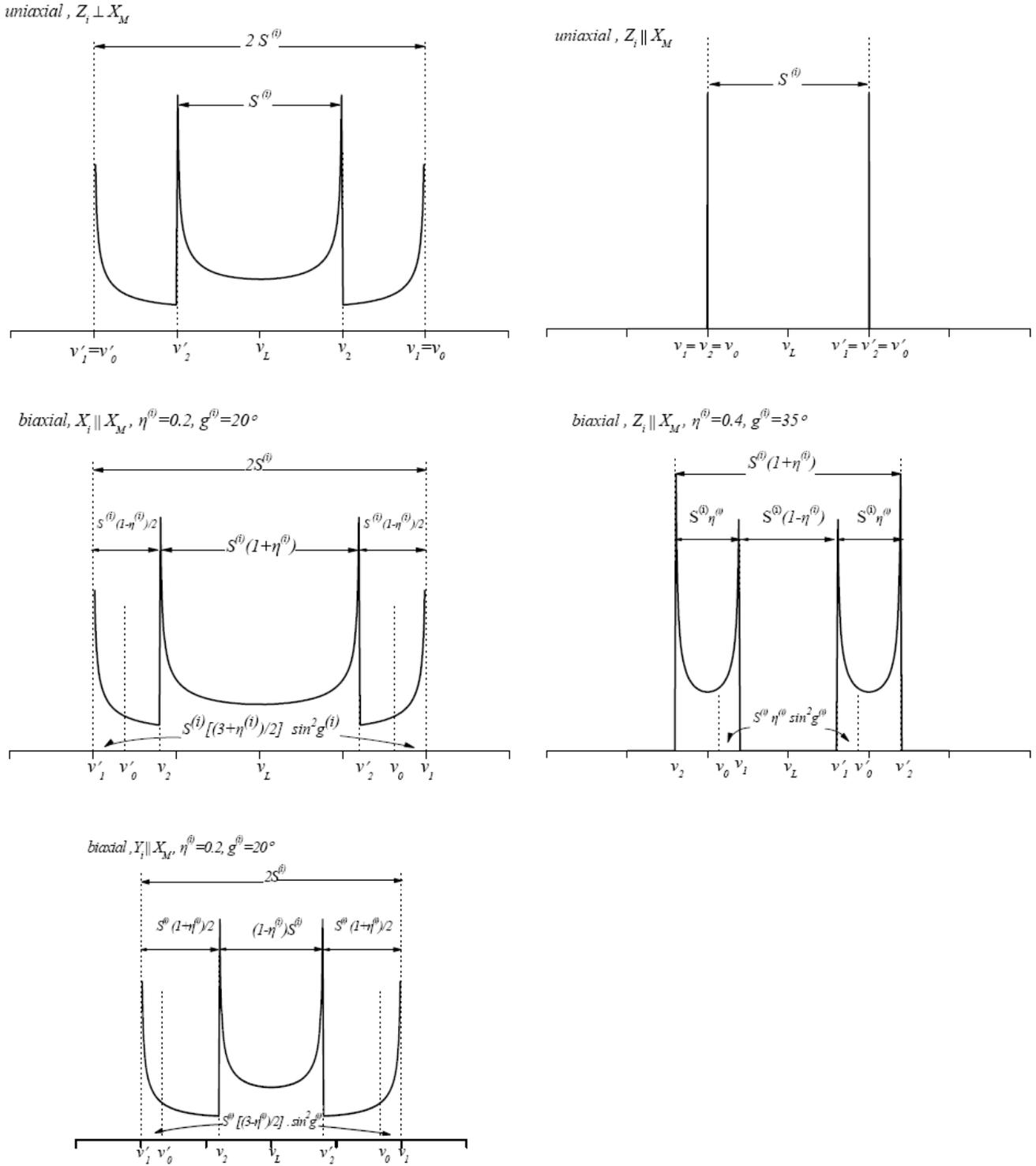

**Figure 2.** Calculated $^2$H-NMR spinning-sample spectral patterns. The spinning axis is perpendicular to the magnetic field and, for all the cases shown, coincides with the axis $X_M$ of minimum diamagnetic susceptibility, which is also taken to be a symmetry axis of the phase. Results are shown for the uniaxial and monoclinic biaxial nematic phases and for all the distinct possibilities of identifying the $X_M$ axis with one of the principal axes $X_i, Y_i, Z_i$ of the field gradient tensor of the particular deuterated site. The Larmor frequency $v_L$ and the frequencies $v_0, v_{0'}$ of the spectral peaks of the magnetically aligned sample ($\theta_M = 0$) are marked on the frequency axis. With the frequencies expressed in units of $3v_Q^{(i)}/4$, the primary order parameter $S^{(i)}$, the biaxiality parameter $\eta^{(i)}$ and the rotation angle $g^{(i)}$ or the dueteriated site are obtained from the spectra as indicated for each of the distinct cases.

Calculated spinning-sample spectral patterns are shown in Figure 2. These are idealized patterns in that they are obtained assuming negligible line broadening, perfectly uniform and planar distribution of the $Z_M$ in the plane perpendicular to the axis of spinning and negligible interference of the spinning frequency with the spectral frequencies *(34, 38)*. The calculated spectra *(20)* correspond to all the possible distinct identifications of the symmetry axis with one of the $X_i, Y_i, Z_i$ in a biaxial phase. The respective spectra of the uniaxial phase are shown for comparison. The same value of the principal order parameter $S^{(i)}$ is used for all the calculated spectra. The following qualitative features of the spinning-sample spectra are apparent from the diagrams of Figure 2: (i) The peaks $\nu_1, \nu_1'$ do not change with symmetry. (ii) With $Z_i$ perpendicular to the spinning axis $X_M$, the peaks $\nu_2, \nu_2'$ are half-way between the centre of the spectrum and the $\nu_1, \nu_1'$ peaks in the case of the uniaxial phase, while for the biaxial phases they move towards $\nu_1, \nu_1'$ or towards the centre of the spectrum, depending on which of the two principal axes $X_i, Y_i$ coincides with $X_M$. (iii) With $Z_i$ parallel to the spinning axis $X_M$, the peaks $\nu_2, \nu_2'$ coincide respectively with the $\nu_1, \nu_1'$ peaks for the uniaxial phase and move further away from the centre of the spectrum in the biaxial phase. Based on these qualitative features one can determine directly from the spinning sample spectrum whether the phase is uniaxial or biaxial and one can also single out which of the $X_i, Y_i, Z_i$ axes coincides with the magnetic principal axis $X_M$. Furthermore, from the positions of the peaks, the parameters $S^{(i)}$ and $\eta^{(i)}$ can be evaluated in each case. However, none of the qualitative or quantitative features of the spinning sample spectrum can differentiate between a monoclinic and an orthorhombic biaxial phase since it contains no information about the frame-rotation angle $g^{(i)}$. The differentiation becomes possible through the comparison of the aligned sample spectrum, whose peaks are at the frequencies $\nu_0, \nu_0'$, with the spinning sample spectrum: For a $D_{2h}$ biaxial phase the aligned sample peaks $\nu_0, \nu_0'$ will coincide in all cases with the peaks $\nu_1, \nu_1'$ while for a $C_{2h}$ biaxial phase they will be positioned between the $\nu_1, \nu_1'$ and the $\nu_2, \nu_2'$ peaks. The value of the angle $g^{(i)}$ is determined directly from the deviation of $\nu_0, \nu_0'$ from $\nu_1, \nu_1'$ as shown on the diagrams. To our knowledge, none of the $^2$H-NMR experiments carried out on biaxial nematic systems where analysed with a view to actually evaluate the angle $g^{(i)}$ from the experimental data; rather, they were analysed by assuming $D_{2h}$ symmetry, i.e. $g^{(i)} = 0$ for all the deuterated sites.

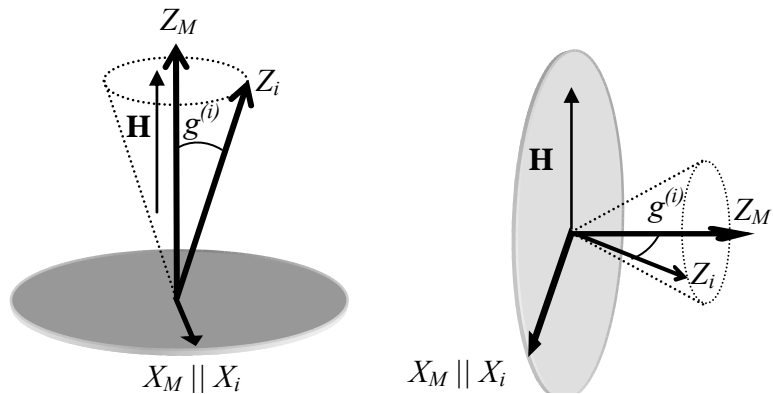

**Figure 3.** Configurations of the principal axes $X_M, Z_M$ and $X_i, Z_i$ relative to the magnetic field *H* for a biaxial nematic phase of monoclinic symmetry in the case where $X_M$ and $X_i$ coincide with the twofold symmetry axis of the phase. (a) Magnetically aligned sample ($\theta_M = 0$), with the $X_M$ axis distributed in the plane perpendicular to *H*. (b) The axis $Z_M$ is held perpendicular to *H* and $X_M$ is distributed in a plane containing *H*. The resulting directional distribution of the $Z_i$ axis is shown for both configurations.

When it is possible to maintain an orientation of the sample with the $Z_M$ axis perpendicular to the magnetic field, as, for example, in lyotropic *(3, 32)* or in polymer nematics *(16)*, the biaxiality of the phase can be inferred from the spectra of the aligned ($\theta_M = 0$) and the perpendicularly rotated ($\theta_M = \pi/2$) sample. This however, cannot differentiate between biaxial symmetries. Furthermore, the values of the primary order parameter and of the biaxiality determined from these spectra will correspond to the true $S^{(i)}$ and $\eta^{(i)}$ only if the symmetry of the phase is $D_{2h}$. The reason is illustrated in Figure 3, where the relative configurations of the $X_M, Z_M$ and $X_i, Z_i$ principal axes, in the case of $C_{2h}$ with $X_i \parallel X_M$, are shown for the aligned and the perpendicularly rotated sample. Clearly, for the aligned sample, the condition $\theta_M = 0$ brings the $Z_i$ axis at a constant angle $g^{(i)}$ with the magnetic field and therefore the splitting $\delta v^{(i)}(\theta_M = 0)$ does not correspond to the maximum value; the latter is obtained when $Z_i$ is parallel to the magnetic field. The direction of magnetic axis $X_M$ could be distributed about the magnetic filed, since the magnetic energy in this case is independent of the azimuthal angle $\phi_M$; however, according to eq(1) this distribution does not affect the splitting $\delta v^{(i)}(\theta_M = 0)$. In the perpendicularly rotated ($\theta_M = \pi/2$) configuration, the axis $Z_i$ is not perpendicular to the magnetic field but forms an angle with it. This angle depends on both, $g^{(i)}$ and the direction of $X_M$. Accordingly, if the direction of $X_M$ is, for any reason, distributed relative to the magnetic field, a corresponding distribution will be generated in the orientations of $Z_i$ as well, in addition to the distribution of the "biaxial" directions $X_i, Y_i$. Therefore, the effects on the resulting spectra will not be restricted to the purely biaxial contributions as would be the case if $g^{(i)}$ vanished, but will include uniaxial contributions originating form the distributed orientations of the principal axis $Z_i$ relative to the magnetic field.

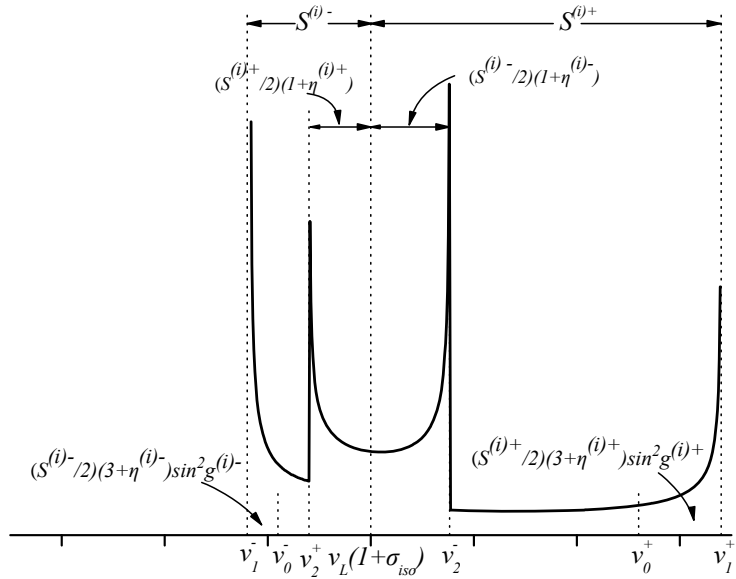

**Figure 4.** Calculated $^2H$-NMR spinning-sample spectral pattern in the biaxial monoclinic nematic phase for a deuterated site exhibiting anisotropic chemical shift asymmetry of relative strength $\lambda^{(i)} = 0.5$ and for the case corresponding to $X_i \parallel X_M$ of figure 2.

The possible existence of substantial *CSA* in a deuterated site leads to aligned and spinning sample spectral patterns with additional structural features, from which the biaxiality parameters associated with both, the $G_{AB}^{(i)}$ and the $C_{AB}^{(i)}$ tensors, can be determined *(20)* and the possibility of $C_{2h}$ phase symmetry can be checked through the evaluation of the respective rotation angles of the principal axis frames of each of these tensors relative to the magnetic principal axes $X_M, Y_M, Z_M$. Representative spectra are shown in Figure 4, where the $S^{(i)\pm}$, $\eta^{(i)\pm}$ and the angles $g^{(i)\pm}$ correspond to the tensors $G_{AB}^{(i)} \pm \lambda^{(i)} C_{AB}^{(i)}$, with the molecular parameter $\lambda^{(i)}$ measuring the strength of the anisotropic part of the *CSA* of the deuterated site *i* in units of the quadrupolar coupling constant $v_Q^{(i)}$.

### 3. Biaxial cluster model of nematics.

None of the presently available experimental results on thermotropic biaxial nematics indicates directly the spontaneous formation of macroscopic biaxial monodomain nematic samples. Rather, all the results are consistent with the presence of local biaxial structures which can be macroscopically ordered into a biaxial nematic state under the action of an external aligning field or of directional surface anchoring. Moreover, some of these results point directly to the presence of orientationally distributed biaxial domains. In particular, the original *NMR* results on biaxiality in bent-core molecules *(9)* can be interpreted equally well in terms of a mono-domain biaxial sample whose transverse diamagnetic anisotropy is uniquely aligned upon spinning in the magnetic field or in terms of a collection of biaxial domains whose transverse axes are randomly distributed until the spinning singles out a common direction for them. On the other hand, the identification of biaxial order in the calamitic tetramer nematics by *NMR (21)* is based directly on the observation of distributed biaxial domains when the aligned sample is rotated to a perpendicular direction relative to the magnetic field. Biaxial order observations by *IR* spectroscopy on the same compounds *(10)*, using surface-aligned samples, showed a nematic-nematic phase transition which, however, did not appear in microcalorimetry measurements on un-aligned samples *(39)*. The *XRD* studies on both types of thermotropic biaxial systems show typical cybotactic cluster diffractograms *(22, 23, 24-25, 40)* analogous to those observed in some conventional uniaxial nematics and therefore do not exclusively imply a biaxial monodomain sample. The electric field response observed by *XRD* in surface aligned samples of bent-core nematics was interpreted as switching of the tranverse axis *(22)* of the biaxial sample. However, electro-optic studies on the same compounds *(14)* support an interpretation of the electric field effects in terms of a field-induced transition from an optically uniaxial to an optically biaxial state, rather than a change in the orientation of the transverse axes in a spontaneously existing biaxial state. Furthermore, the presence of biaxial clusters is invoked for the interpretation of the observed field -induced texture transitions in nematic dimers containing bent-core units *(41)*. These considerations, together with the broader relevance of local structures in liquid crystal phase transitions, and indeed in liquid-liquid transitions *(42)*, have motivated the development of the cluster model of biaxial nematics *(17)*. The essential element of this model is a nematic phase consisting of biaxial microdomains which, in the absence of an external aligning stimulus, are randomly distributed into a macroscopically uniaxial nematic state. Transitions from this state to a biaxial state, either field-induced or driven by decreasing temperature, are the result of the collective alignment of the microdomains along a common direction transverse to the uniaxial director. In this sense the transitions are termed as poly-domain to mono-domain phase transitions. Uniaxial-uniaxial nematic transitions are also present in this model, where the two uniaxial phases differ with respect to the size and internal biaxial order of the clusters.

In the simplest phenomenological formulation of the cluster model *(17)*, a nematic sample with perfect molecular alignment along the longitudinal direction **n** is divided into biaxial clusters (Figure 5a). An order parameter *σ* describes the average internal biaxial order of the clusters and another order parameter *q* describes the overall biaxial order in the macroscopic sample. The free energy of this system includes (i) the internal energy and entropy of the individual clusters, (ii) the interaction energy among neighboring clusters and (iii) the collective entropy associated with the relative magnitudes and orientations of the clusters. The Landau-de Gennes (*L-dG*) expansion of the free energy in terms of the order parameters *σ* and *q*, including the orientational coupling of the biaxial clusters to an applied transverse electric field of strength *E*, is of the form:

$$F(T;q, \sigma) = F_0(T) + F'(T,\sigma) + F''(q^2) - e\sigma q^2 - hE^2 q \qquad , \qquad (2)$$

where $e$ and $h$ are positive coupling constants, $F''$ contains at least up to fourth power terms in $q$ and $F'$ contains at least up to third power terms in $\sigma$. With $F'$, $F''$ restricted to their minimal-power form *(17)* and with the temperature dependence carried exclusively by the first-power term of $F'$, the system, in the absence of an external field, can exist in three distinct phases (see Figure 5a) (i) a macroscopically biaxial phase $N_b$, with $q \neq 0, \sigma \neq 0$, (ii) a macroscopically uniaxial phase $N_{u'}$ of biaxial clusters, with $q = 0, \sigma \neq 0$ and (iii) a proper uniaxial state $N_u$ with $q = 0, \sigma = 0$. Depending on the relative values of the expansion parameters, the model can accommodate different phase sequences on decreasing the temperature from the $N_u$ phase down to the tempereture where a positionally ordered phase denoted by $X$ (smectic, columnar or solid) sets in. These sequences are shown in Figure 5b. Possible direct transitions from the isotropic to any one of the $N_{u'}$, $N_b$ or $X$ phases, being irrelevant to the present discussion, are not included in the figure.

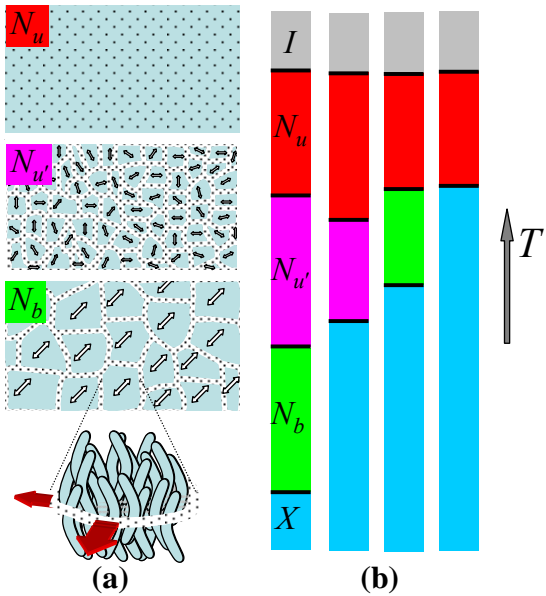

**Figure 5.** (a) Cross section of a nematic sample, with the axis of perfect molecular alignment directed perpendicular to the plane of the figure, illustrating its biaxial cluster composition in the three possible phases of the model system. (b) Schematic representation of possible thermotropic phase sequences of the system.

The two uniaxial phases may differ by orders of magnitude in their response to the applied electric field *(17)*. In the $N_u$ phase, the electric field addresses essentially the individual molecules and therefore, at the temperatures and field strengths of practical interest, it has marginal effects on the biaxial ordering of the sample. Specifically, it gives rise to a para-biaxial nematic which, at very high field strength, undergoes a first-order transition to the $N_b$ phase. In contrast, the $N_{u'}$ acquires substantial field-induced biaxial order at relatively low electric field strength. In this case the field addresses entire biaxial clusters and the critical field values for the transition from the para-biaxial to the $N_b$ phase can be lower by one or two orders of magnitude compared to the respective values for $N_u$. Stated in slightly different terms, the application of an electric field elevates the temperature of the $N_{u'}$ - $N_b$ phase transition much more than it elevates the respective temperature for the $N_u$ - $N_b$ transition. In this respect, the observed thermotropic biaxiality in the bent-core and in the calamitic tetramer systems, if entirely field-induced, would place them in the phase sequence category of the second row in Figure 5b, i.e. the $I$ - $N_u$ - $N_{u'}$ - $X$ sequence, with the $N_b$ appearing as a stable phase only on applying a field. Accordingly, the common uniaxial nematics would correspond to the $I$ - $N_u$ - $X$ (fourth row in Figure 5b).

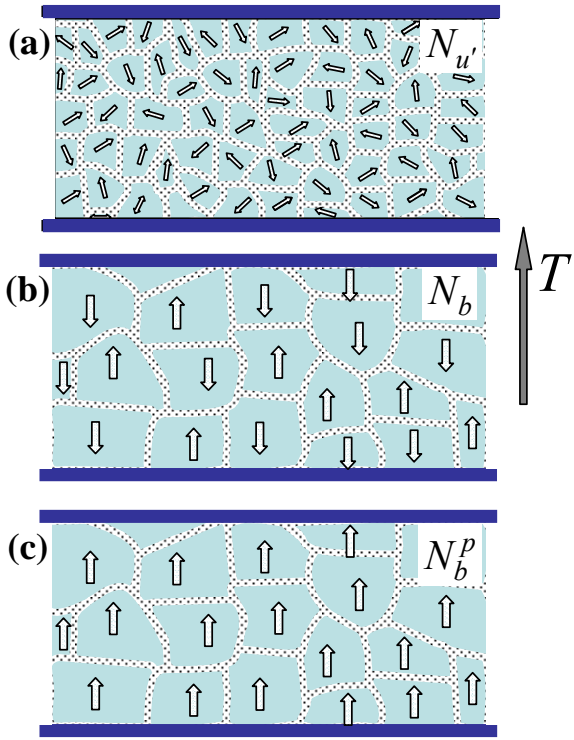

**Figure 6.** Cross section of a model nematic sample consisting of polar clusters. The directional disposition of the clusters is illustrated for (a) the macroscopically uniaxial phase $N_{u'}$, (b) the biaxial apolar phase $N_b$ and (c) the biaxial polar phase $N_b^p$.

The cluster model is readily extended to include the possibility of local biaxial and transverse polar order. The latter is directly relevant to bent-core compounds in view of their well known spontaneously polar self-organisation in the smectic phases *(23)* and of the identification of local polar ordering in atomistic simulations *(43)*. Clearly, assuming that the clusters in the sample are polar in the transverse directions, automatically renders them biaxial as well. In this case the biaxiality of the clusters is generated entirely by their polarity and the primary order parameter $\rho$ of the clusters is defined as the average magnitude squared of the cluster polar vector. The macroscopic polar order is described by the average polarity vector $\vec{p}$ of the entire sample. Thus, the basic order parameters for this system are three: the macroscopic biaxiality order parameter $q$, the macroscopic polarity order parameter $p \equiv |\vec{p}|$ and the cluster polarity order parameter $\rho$. The *L-dG* formulation of the free energy in this case includes in eq (2) the parameter $\rho$ in place of $\sigma$, and the additional terms $-e'\rho p^2, -e''qp^2$ and $-h'(\vec{E} \cdot \vec{p})$. Here, $e', e'', h'$ are positive constants describing the coupling of the macroscopic polar order to the polar order of the clusters, to the macroscopic biaxial order and to the applied electric field, respectively. The possible nematic phases for this system, in the absence of an external field, are four: a proper uniaxial $N_u$ phase ($\rho = q = p = 0$), a uniaxial $N_{u'}$ phase with local polar order ($\rho \neq 0, q = p = 0$), a biaxial apolar $N_b$ phase ($\rho \neq 0, q \neq 0, p = 0$) and a biaxial polar phase $N_b^p$, with $\rho \neq 0, q \neq 0, p \neq 0$. The cluster compositions of the $N_{u'}$, $N_b$ and $N_b^p$ are depicted in Figure 6. The presence of the polar $N_b^p$ phase below the $N_b$ phase in the temperature scale, extends the possible phase sequences of Figure 5b accordingly and includes, among othr possibilities, direct $N_{u'}$ - $N_b^p$ phase transitions.

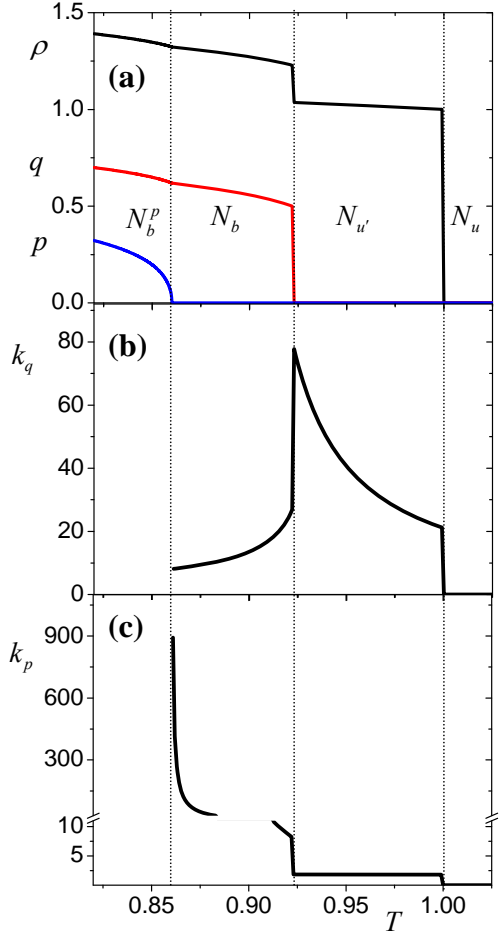

**Figure 7.** (a) Calculated temperature dependence of the order parameters $\rho, q, p$ (scaled values) for the full sequence of possible nematic phases of a model system forming polar clusters. The reduced temperature scale is arbitrary. (b) The corresponding temperature dependence of the electro-biaxial coefficient $k_q$ and (c) of the electro-polar coefficient $k_p$ (scaled values, in arbitrary units).

The diagrams of Figure 7a show the variation of the order parameters $\rho$, $q$, $p$ with temperature for a representative combination of the *L-dG* expansion parameters allowing the full sequence of the possible nematic phases. The diagrams of Figure 7b describe the respective response of these nematic phases to a week electric field. Due to the transverse polarity of the clusters, the system has two modes of response to the electric field, a linear one, corresponding the coupling of the polar order to the field, and a quadratic, corresponding to the coupling of the biaxial order. The two modes of response are quantified by means of the electro-polar coefficient $k_p \equiv (\partial p / \partial E)_{E \to 0}$ and the electro-biaxial coefficient $k_q \equiv (\partial q / \partial E^2)_{E \to 0}$. These modes can be of comparable magnitudes in the $N_{u'}$, which would lead to unusual electro-optic behaviour of this macroscopically uniaxial and apolar nematic phase.

The results presented here are based on a primitive formulation of the model *(17)*, with the primary orientational order of the molecules frozen to its maximum and ignoring all but the minimal power terms in the *L-dG* expansion of equation (2). More general and more realistic formulations of the phenomenology have been studied, extending the implications of the cluster model to the isotropic-uniaxial nematic phase transition. The results are presented elsewhere *(44)*. Finally, to identify reliably local structures, computer simulations based on very large samples in the nematic phase are required. Such studies are presently under way for molecular models of bent-core systems. Preliminary results support the presence of sizable clusters whose internal ordering is of lower symmetry than that of the bulk sample.

**4. Discussion and conclusions**

In this work we have addressed two fundamental aspects of biaxial nematic order: its symmetry and its spatial uniformity. The explicit consideration of the full range of possible symmetries for the biaxial nematic phase was shown to be necessary for the consistent analysis of the experimental measurements of phase biaxiality in the new thermotropic biaxial nematics. Naturally, such measurements concern physical properties that are sensitive to biaxial orientational order. However, not all of these measurements are equally sensitive in differentiting between the $D_{2h}$ symmetry and the lower symmetries, $C_{2h}$ and $C_i$. In other words, while nematic biaxiality can be demonstrated by measuring a single tensor quantity, a reliable $D_{2h}$ symmetry assignment would require the measurement of several different tensor quantities to ensure that all the principal axes are commonly directed along three orthogonal directions. As shown in Section 2, the $^2$*H-NMR* methodology can be very efficient in identifying deviations from the $D_{2h}$ symmetry. Other methods, such as *IR* spectroscopy, which are also based on the measurement of the orientational order of specifically labeled molecular segments, can also be sensitive to the particular symmetry of the biaxial ordering. On the other hand, anisotropic bulk properties, particularly optical, dielectric and magnetic, are known from the conventional calamitic smectic phases not to be sensitive indicators of the symmetry differences as they do not show large relative deviations of their principal directions in the $C_{2h}$ biaxial phase. However, this insensitivity would not necessarily persist in bent – core nematics or in other compounds with more complex molecular architectures than the conventional calamitic or discotic mesogens. In that case, a biaxial $C_{2h}$ symmetry would entail, for example, substantial deviations of the optical principal axes from the directions of the dielectric or of the magnetic principal axes.

The spatial uniformity aspect of the biaxial orientational order in nematics was addressed in the context of the biaxial cluster model. Therein, the establishment of macroscopic biaxial order is the result of a poly-domain to mono-domain phase transition in a system consisting of biaxial clusters that are uniaxially distributed in the poly-domain state. This transition can be thermotropic or field induced. Within this model, a number of recent experimental observations on thermotropic nematic phase biaxiality can be rationalized and interpreted consistently. These include observations of the intriguing electro-optic behaviour of bent-core nematics *(14, 41, 45-46)*, biaxiality measurements by $^2$*H-NMR* in bent-core and in calamitic-tetramer nematics, local biaxial structures revealed in *XRD* of a variety of bent-core and laterally substituted calamitic nematics *(22-25, 40)* and phase biaxiality measurements by *IR* sperctroscopy in calamitic tetramers *(10)* in relation to the results of calorimetry studies *(39)* on the same compounds. Furthermore, in the context of the cluster model, the above observations can be naturally related to direct observations of domain structures in bent-core nematics *(18)* and to the results of previous studies which invoked the presence of local structures in order to interpret the behavior of bent-core uniaxial nematics *(47)*.

The emerging picture of the thermotropic biaxial nematics, which combines the different symmetry possibilities with the biaxial cluster composition of the $N_{u'}$ phase, widens dramatically the scope of the phenomenological description of the nematic phase; it obviously includes the conventional, $D_{2h}$ and spatially uniform, model as special case. On the other hand, this wider picture has significant implications on the alignment of biaxial nematics: Due to the micro-domain structure, the application of an external field induces a biaxial, and uniformly aligned, state of the sample rather that reorienting a spontaneously existing biaxial sample. Naturally, the alignment via a field-induced transition involves a critical value of the field, below which the system appears not to be susceptible to alignment. Moreover, in the case of sufficiently strong electro-polar response of the clusters, the alignment can show a second, polar, stage on increasing the external field. Aside from that, the aligning effect caused by the electric field on the different anisotropic properties of the sample will depend on the symmetry of the biaxial order. Thus, for $D_{2h}$ symmetry all the anisotropic properties will have one of their principal axes aligned in the direction of the field. For $C_{2h}$, such alignment will result only if the principal axis of largest dielectric constant coincides with the twofold symmetry axis.

**Acknowledgement**

This work was funded by the EC through the project BIND-216025 (Biaxial Nematic Devices), FP7 / ICT-1-3.2 / STREP-CP-FP-INFSO)**References**

(1) Freiser, M. J. *Phys. Rev. Lett.* **1970**, *24*, 1041-1043.

(2) Berardi, R.; Muccioli, L.; Orlandi, S.; Ricci, M.; Zannoni, C. *J. Phys.: Condens. Matter.* **2008,** *20*, 463101 and references therein.

(3) Yu, L. J.; Saupe, A. *Phys. Rev. Lett.* **1980**, *45*, 1000-1003.

(4) Leube, H. F.; Finkelmann, H. *Makromol. Chem.* **1991**, *192*, 1317-1329.

(5) Praefcke, K.; Blunk, D.; Singer, D.; Goodby, J. W.; Toyne, K. J.; Hird, M.; Styring, P.; Norbert, W. D. J. A. *Mol. Cryst. Liq. Cryst*. **1998**, *323*, 231-259.

(6) Kouwer, P. H. J.; Mehl, G. H. *J. Am. Chem. Soc.* **2003**, *125*, 11172-11173.

(7) Date, R. W.; Bruce, D. W. *J. Am. Chem. Soc.* **2003**, *125*, 9012-9013.

(8) Dingemans, T. J.; Samulski, E. T. *Liq. Cryst.* **2000,** *27*, 131-136.

(9) Madsen, L.A.; Dingemans, T.J.; Nakata, M.; Samulski, E.T. *Phys. Rev. Lett.* **2004**, *92*, 145505.

(10) Merkel, K.; Kocot, A.; Vij, J. K.; Korlacki, R.; Mehl, G. H.; Meyer, T. *Phys. Rev. Lett.* **2004**, *93*, 237801.

(11) Luckhurst, G. R. *Nature* **2004,** *430*, 413-414.

(12) Apreutesei, D.; Mehl, G. H. *Chem. Commun.* **2006,** *6*, 609-611.

(13) Cuetos, A.; Galindo, A.; Jackson, G. *Phys. Rev. Lett.* **2008,** *101*, 237802.

(14) Lee, J.-H.; Lim, T. K.; Kim, W. T.; Jin, J. I. *J. Appl. Phys.* **2007,** *101*, 034105.

(15) Berardi, R.; Muccioli, L.; Zannoni, C. *J. Chem. Phys.* **2008,** *128*, 024905.

(16) Severing, K.; Saalwachter, K. *Phys. Rev. Lett.* **2004,** *92*, 125501.

(17) Vanakaras, A. G.; Photinos, D. J. *J. Chem. Phys.* **2008,** *128*, 154512.

(18) Görtz, V.; Southern, C.; Roberts, N.W.; Gleeson, H. F.; Goodby, J. W. *Soft Matter*. **2009,** *5*, 463-471.

(19) Shimbo, Y.; Gorecka, E.; Pociecha, D.; Araoka, F.; Goto, M.; Takanishi, Y.; Ishikawa, K.; Mieczkowski, J.; Gomola, K.; Takezoe, H. *Phys. Rev. Lett.* **2006,** *97*, 113901.

(20) Karahaliou, P. K.; Vanakaras, A. G.; Photinos, D. J. (2009)

(21) Figueirinhas, J. L.; Cruz, C.; Filip, D.; Feio, G.; Ribeiro, A. C.; Frère, Y.; Meyer, T.; Mehl, G. H. *Phys. Rev. Lett.* **2005,** *94*, 107802.

(22) Acharya, B. R.; Primak, A.; Kumar, S. *Phys. Rev. Lett.* **2004,** *92,* 145506.

(23) Takezoe, H.; Takanishi, Y. *Jpn. J. Appl. Phys.* **2006,** *45*, 597-625 and references therein.

(24) Karahaliou, P. K. ; Kouwer, P. H. J.; Meyer, T.; Mehl, G. H.; Photinos, D. J. *Soft Matter*. **2007,** *3*, 857-865.

(25) Karahaliou, P. K.; Kouwer, P. H. J.; Meyer, T.; Mehl, G. H.; Photinos, D. J. *J. Phys. Chem. B*. **2008,** *112*, 6550-6556.

(26) Mettout, B. *Phys. Rev. E*. **2006,** *74*, 41701.

(27) Brand, H. R.; Cladis P. E.; Pleiner, H. *Int. J. Eng. Sci.* **2000,** *38*, 1099-1112.

(28) Dong, R. Y., *Nuclear Magnetic Resonance of Liquid Crystals, 2nd ed.*; Springer-Verlag, 1997.14